# Identifying and analysing toxic actors and communities on Facebook by employing network analysis

R. Manuvie[1], S. Chatterjee[2]

[1]*University College Groningen, University of Groningen, r.manuvie@rug.nl (corresponding author)*

[2]*Foundation The London Story, Fluwelen Burgwal 58, 2511CJ, Den Haag, The Netherlands, saikat@thelondonstory.org*

## Abstract

There has been an increasingly widespread agreement among both academic circles and the general public that the Social Media Platforms (SMPs) play a central role in the dissemination of harmful and negative sentiment content in a coordinated manner. A substantial body of recent scholarly research has demonstrated the ways in which hateful content, political propaganda, and targeted messaging on SMPs have contributed to serious real-world consequences. Adopting inspirations from graph theory, in this paper we apply novel network and community finding algorithms over a representative Facebook dataset (n=608,417) which we have scrapped through 630 pages. By applying Girvan-Newman algorithm over the historical dataset our analysis finds five communities of coordinated networks of actors, within the contexts of Indian far-right Hindutva discourse. This work further paves the path for future potentials of applying such novel network analysis algorithms to SMPs, in order to automatically identify toxic coordinated communities and sub-communities, and to possibly resist real-world threats emerging from information dissemination in the SMPs.

## 1. Introduction

Studies show that disruptive actors around the world have been successfully using Social Media Platforms (SMPs) to fuel hate speech against minorities and spread negative sentiment contents (see Belew and Massanari 2018, Matamoros-Fernández and Farkas 2021, Manuvie and Chatterjee 2023). Consequently, these actors tend to manipulate the public discourse by suppressing independent voices through counter-speech, or excessive content production which disrupts the core values of democratic structures, often acting in a coordinated way. In the present study, our goal is to apply novel community searching algorithms over a select Facebook dataset of 630 pages and algorithmically analyse their networked and coordinated behaviour in the digital domain. These pages were identified on Facebook using the keyword search methods and following Facebook's recommendation system between April 2020 and December 2020. Keywords surrounding popular discourses such as "Love Jihad", "Corona Jihad", and "spit-jihad" alongside identity-specific slang terminology were used to identify the initial set of pages. Subsequently, the Facebook recommendation system was used to manually 'like' and collect URLs of 803 pages that are associated with cross-sharing and cross-posting anti-muslim and anti-migrant content and discourses respectively. Through careful human iteration, the size of the list of pages was reduced to 653 pages with page likes above 1000 likes, and a posting frequency of daily. As the access to the CrowdTangle platform was made available to us, these pages were batch uploaded to the CrowdTangle dashboard. Of the total 653 pages, only 630 Facebook pages were tracked by CrowdTangle due to system limitations in terms of privacy and performance of the page at the time of the upload. The 630 pages were further sorted based on discourses, narratives, and fan-following of individual political leaders or ideological groups into CT lists.

The pages were further monitored for qualitative analysis to ensure that these pages fall in the category of regular hate-speakers or perpetrators of hateful content. Results of these analyses were shared with Facebook's parent company Meta, focusing on extremely hateful content and lists. In response to our report, Meta provided a statement suggesting that it deploys an automated language processing system to flag and remove content that is violative of its hate-speech policies, or where a post falls into the category of a dangerous organisation or dangerous actor. We, therefore, decided to run two open access sentiment analysis and hate-speech detection models over our CrowdTangle dataset of toxic actors (see Manuvie and Chatterjee, 2023). Textual content associated with each post along with the predicted sentiment and hate-speech labels is released in our public GitHub page.[1] These sequences of text strings and associated sentiment and hate speech labels (as predicted by the XLM-T-based language models) can be considered as "weak-labelled" datasets for further research purposes. Backed by our qualitative discourse analysis of these groups over the last two years, we can confirm that not everything that is posted by these actors can qualify to be hate-speech content, or can be validly detected. However, our results establish that the amount of negative messaging and hate content posted on these groups repeatedly should be sufficient to de-platform at least some of them under Meta's current policies on content moderation.

The goal of the present paper is to further extend our previous study and to exhibit how community finding algorithms can be applied on such datasets to find underlying network effects, and to (algorithmically) find underlying communities or groups of actors that are responsible for disseminating information in the SMPs in a coordinated manner. To do this, we have structured the paper in the following way. In section 2, we discuss the methodology underlying our data selection and data archiving process. In section 3, overall statistics of the dataset is shown. The detailed processes of data-reduction and network analysis are subsequently presented in section 4.

## 2. Methodology

The overarching goal of our research is to search, identify and validate Facebook actors (i.e., pages or groups) that are part of an information ecosystem – which allegedly spreads hate speech and disinformation. Such eco-systems of information dissemination and circulation can potentially form "echo chambers" connected to a certain ideology or narrative or even can create social polarization on digital platforms. We study the patterns in the behaviour of the actors within the Facebook platform (i.e., link sharing) and further attempt to identify the groups and communities of information dissemineation by employing network analysis. The first step to identify the toxic actors is to begin with a primary list of toxic fan pages and groups that contribute to the process of disseminating hate narratives against religious minorities in India. To do this, we chose the following as the determinant factors*:*

---

[1]

a)  The individuals who have a public track record on hate speech and/or providing hate narratives targeting religious minorities particularly, Muslims.
b)   They have declared or exhibited their affinity toward the Hindutva ideology.

The individuals are chosen due to their influential position in society to mobilise people against the minorities in the pursuit of achieving the Hindutva aims. We also document their nexus to governance institutions, political leadership and Hindutva-based organisations. Here, we use the following indicators to record their influential position in society - politicians, media persons, vigilante leaders, Hindutva-based religious influencers, state officials, or Government officers. Built on this selection process, the second task for mapping and monitoring the fan pages and the groups generating hate content in the name of hate actors (individuals) was, to list the fan pages and groups found through a simple search on Facebook and CrowdTangle platforms with their names. CrowdTangle Search tool helps in searching content across social media by putting a keyword, hashtag or URL into the search bar. We modified the search tool to include filters to sort by countries, language options (Hindi and English) and timeframe. Based on such search combinations, we found an initial list having 630 pages spreading hate narratives.

However, the number of pages fluctuates, as the pages are removed (by Facebook), added (as the new pages are formed), or deleted from Facebook (by page admins or by Facebook). Although such fluctuations do affect the overall statistics, in our research process, we download the historical dataset (in CSV format) and archive this dataset with a time-stamp, so that our findings/claims can be reproduced later using that archived dataset – even if some pages or groups from our compiled lists are deleted later from the Facebook platform. Subsequently, for data visualisation and inspecting, we ingest the dataset to an elasticsearch database and plot various visualizations using Kibana frontend (hosted by Elastic Cloud).[2]

### 3.   Dataset and the overall statistics

Corresponding to each entry of the CrowdTangle data, there are 40 metadata fields as shown in Appendix 1. Following the CrowdTangle data sharing policies, we can not publicly share the contents of the posts. So, in the next subsections we only show the results derived by processing and analysing the dataset that is scrapped using CrowdTangle. As of 17th of  April, 2023, we have scraped the Facebook historical data corresponding to 630 Facebook pages using the CrowdTangle platform from 31-12-2014 onwards and have performed our statistical analysis in this paper. The dataset consists of a total 608,417 entries. In order to inspect the timeline of the posts, we use the 'Post Created Date' and 'Post Created Time' fields. As shown in Appendix 1, the integer values corresponding to the 'statistics' object provide us with information about likes, comments, shares counts. The URL strings corresponding to the field 'expandedLinks.original' provide us with the statistics of the most shared links, which we have compiled in Appendix 1. Text entries corresponding to the fields like 'Page Description', 'message', 'image text' or 'title' are used to understand the overall narratives and for plotting word clouds.

---

[2] To see the live dashboard, visit this [link](link).

In order to have a broader overview of the activities of the actors, we begin with a time-histogram of the datasets. In Figure 1, we show the histogram of the time series data that is scrapped between the dates 31-12-2014 and 17-04-2023. The top 20 actors are shown in the pie diagram of Figure 2. To provide a more intuitive visualisation about the relative activity of the actors, top 50 actors are shown in Figure 3 in an word cloud. Concerning the distribution of the types of posts, 45.14%, 32.48%, 10.9%, 7.12% and 3.02% contents correspond to photos, links, native video, status and live videos respectively. Considering the top admin countries, 88.06% of posts correspond to India (IN), 10.5% admins are from South Africa (SA), 0.6% from Australia (AU), 0.47% from Pakistan (PK) and 0.25% from Bangladesh (BD). Other admin countries present in the dataset are, the United States (US), Canada (CA) and Bhutan (BT). We have also extracted and plotted the distribution of branded content sponsors in Figure 4. Moreover, in order to understand the overall narrative and sentiment of messages, we have also mapped a word cloud of the shared messages in Figure 5. The plot shows their evident leaning towards right-wing Hindutva sentiment through the most occurring phrases within the word cloud, like *Jay Shree Ram*, *Bajrang Dal*, *Kattar Hindustani*, *Jai Hind* etc.

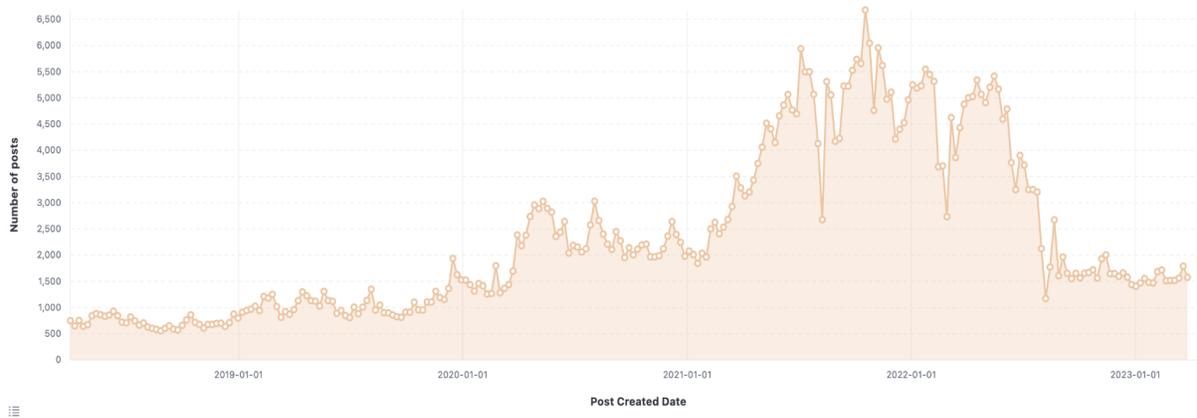

Figure 1. Time histogram of the number of posts.

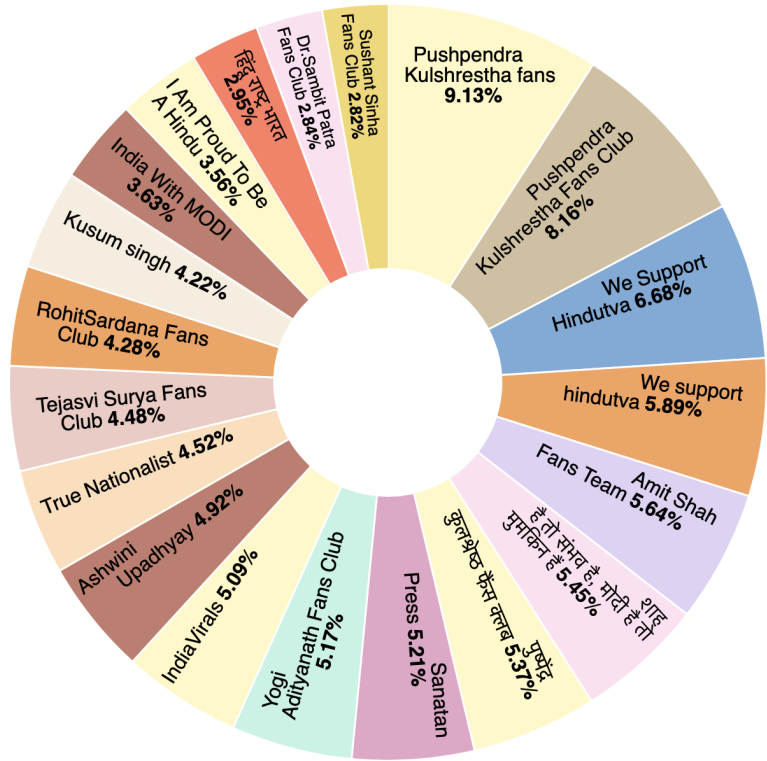

Figure 2. Top 20 actors and their percentage of posts within the dataset.

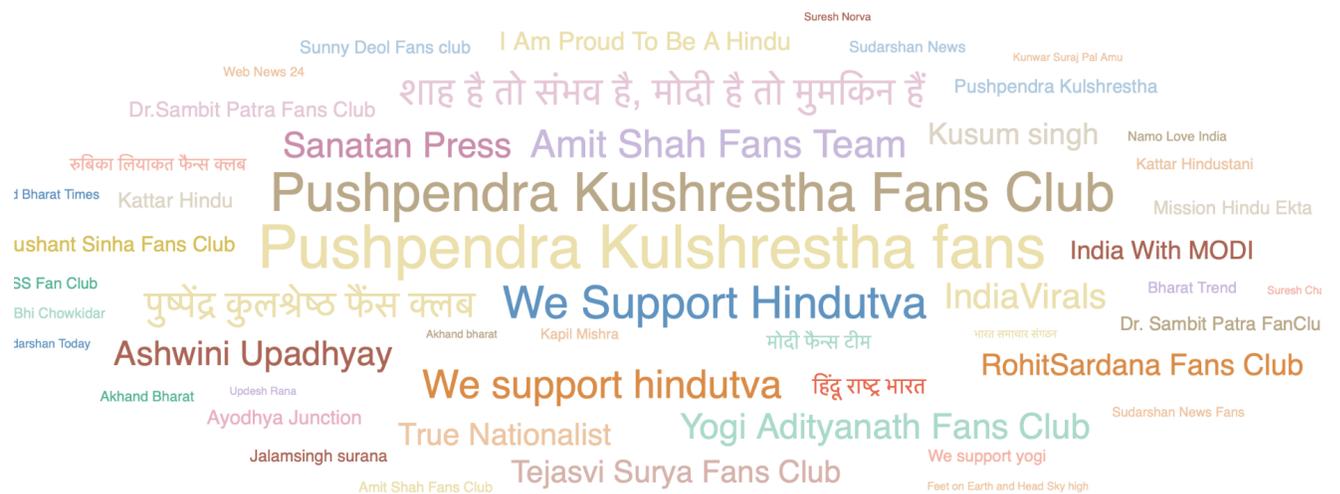

Figure 3. Wordcloud of the top 50 actors.

Figure 4. Percentages of branded content sponsors within the dataset.

Figure 5. Wordcloud of messages shared by the actors.

In Appendix 2, we show the 100 most shared links and their corresponding counts. We have also implemented an automated URL validator which checks whether the links are valid, or they return a 404 error.[3] We have run this algorithm on our dataset and hereby report that 57.3% of the top 1000 non-Facebook links are identified to be invalid or broken. This implies that the contents corresponding to these links are deleted by the owner from their site. It is important to note here that when Facebook contents are taken down either by the page/group admin or by Facebook, the URLs don't return us a 404 error. So, through the present version of our URL validator algorithm, it is not possible to identify such webpages. Hence, we only applied our algorithm to the top 1000 non-Facebook links (i.e., we filtered out all the links from our list which contain "www.facebook.com").

## 4. Network analysis

Networks are commonly described by sets of two items, "nodes" and "edges" – which are the basic building blocks of a mathematical *graph*. Edges are connective lines between two nodes. In order to visualise the graph of link-sharing behaviour and connectivities between various actors in our dataset, we have mapped the whole dataset into a graph; subsequently performed graph network analysis; and finally visualised the identified sub-graphs and communities into Gephi. Methodologically, this is done by creating separate nodes and edges data frames. The nodes have the map (**Id, Label**), whereas the edges have the map of (**Source, Target**). We use the entries of '*account.name*' field as our 'Source', and '*expandedLinks.original*' as our 'Target'. We thereafter employ the Python Networkx package to perform novel network analysis in order to analyse the graph to find suspected coordinated behaviours and communities. Finally we also create network visualizations of the detected communities by employing the visualisation software Gephi[4]. The network maps together with graph analysis help us to understand the connectivities between different groups of actors and furthermore show how these networks of information eco-systems are connected through link-sharing behaviour.

### Graph centrality analysis

In order to characterize the importance of different nodes within the network, we compute the ***degrees*** corresponding to each node. The degree is a measure that provides us with the number of edges each node has. In other words, the degree tells us how many neighbours a particular node has. It can be assumed that nodes which have the most edges are the most important or central within the network as they are directly connected to lots of other nodes. Hence, nodes with high degrees are expected to be important actors in the network. In other words, nodes with a high degree tend to be more influential and popular in the network. These individuals may be well-connected to many others and have a greater ability to spread information or ideas. The degree of nodes can also help us understand the clustering of the network - that is, how tightly connected groups of nodes are to one another. If nodes in a group have a high degree of connection to each other but low connections to nodes outside the group, it indicates a tightly-knit community. Within our data, which contains a total of **414,490 nodes** and **537,225 edges**, the maximum degree of the graph is 23497 associated with the page "*The Kapil Sharma Fan Club*". We also compute the average degree of the nodes in the network, which is 2.6.

---

[3] See our ipython notebook code for the automated URL validation: https://github.com/LondonStory/CrowdTangle-New-Actor-Searching-Algorithm/blob/main/Automated_URL_validator.ipynb.

[4] The ipython Google Colab notebooks that perform network analysis and prepares the nodes and edges files for Gephi visulaisation can be found in our public GitHub page at https://github.com/LondonStory/CrowdTangle-Network-Analysis.

The network graph created above by mapping the actors and their shared links, contains **54 *connected components***. The number of connected components in a graph is a way of measuring how many separate clusters or groups of nodes exist in the graph. A connected component is a subgraph of a graph in which every node is connected to every other node by at least one path. In other words, all nodes in a connected component can be reached from any other node in the same component by traversing edges in the graph. Presence of 54 connected components in our dataset implies that the whole network is not fully connected and there are distinct communities or groups of nodes.

Another associated network centrality measure that we computed next is *degree centrality*. This is one of the metrics that is used in the literature to evaluate the importance of a node and is defined by the number of neighbours that a node has divided by the total number of neighbours that the node could possibly have:

***Degree centrality*** *= No. of neighbours of a node / No. of neighbours the node could possibly have*

Because in social networks self-loops are not allowed (i.e, an actor can't follow itself), the number of neighbours an individual actor can possibly have, counts to every other node within the network, excluding itself. We find that the maximum degree of centrality of our network corresponds to the page "Pushpendra Kulshrestha Fans Club", with a value of 0.057. In order to understand the correlation between the degrees and the degree centrality measures of the nodes, in Figure 6 we plot a scatter diagram, showing the distribution of degrees in the horizontal axis versus the degree centralities of the nodes in the vertical axis. In Table 1 we also show the top 10 actors who have the highest degree centrality within the dataset. The scatter plot between degree centrality and degrees is employed in order to understand the relationship between these two measures. Nodes with higher degree centrality and degree are typically considered more important in the network, and considered to play key roles in information flow. Usually, nodes with higher degrees tend to have higher degree centrality, which indicates that they are more important within the network. However, there may be outliers or 'anomalous' nodes within the network, where nodes have a high degree centrality despite having a relatively low degree. These nodes may act as 'hubs' or bridges between different parts of the network.

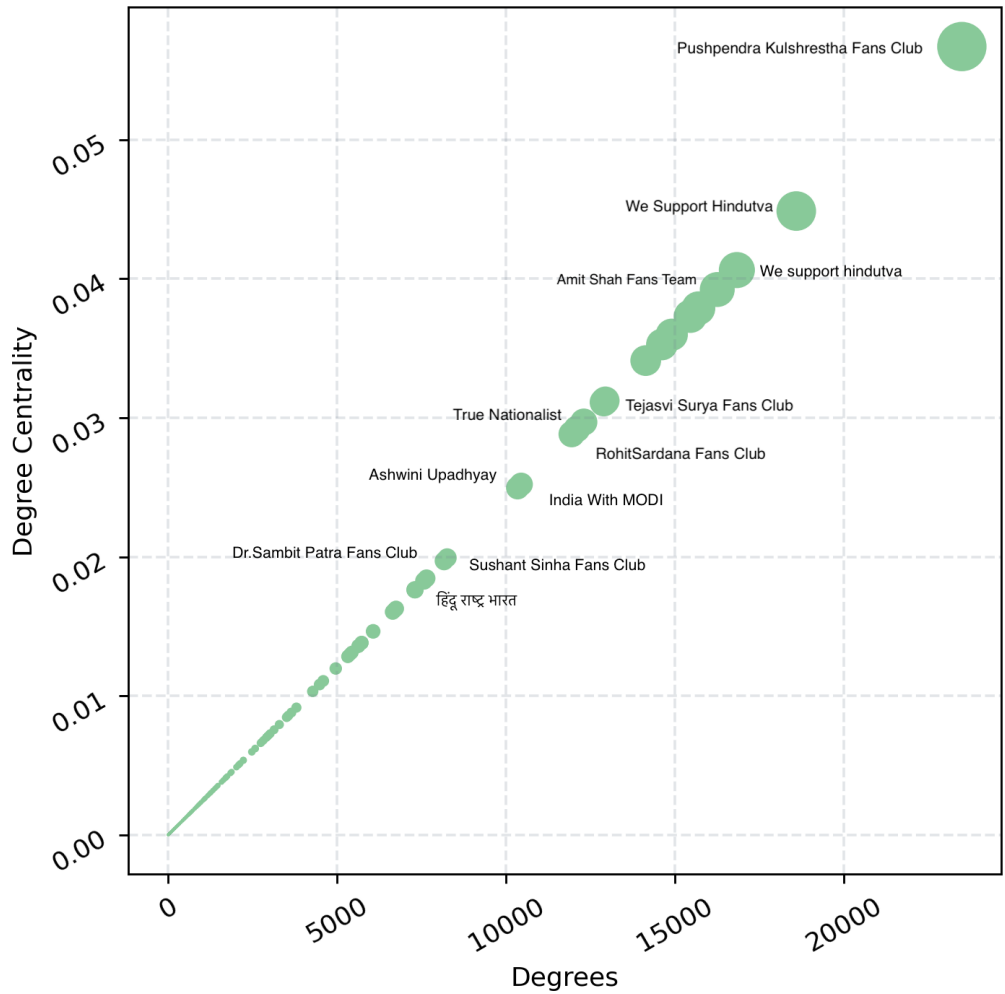

Figure 6. Scatter plot between degrees vs degree centrality measures of the nodes.

By analysing Figure 6 and Table 1, we see that there's a linear correlation between the degrees and degree centrality measures. Top two actors (namely, *Pushpendra Kulshrestha Fans Club* and *We Support Hindutva*) generally surpass the other actors regarding their engagement in the network. Top 3rd – 9th actors from Table 1 form a cluster in the scatter plot. The next cluster we identify consists of *Tejasvi Surya Fans Club*, *True Nationalist* and *RohitSardana Fans Club*. Other smaller clusters are highlighted in Figure 6.

**Table 1. Top 10 pages with highest degree centrality measures in the whole graph**

| Page name | Degree centrality |
|---|---|
| Pushpendra Kulshrestha Fans Club | 0.057 |
| We Support Hindutva | 0.045 |
| We support hindutva | 0.041 |
| Amit Shah Fans Team | 0.039 |
| शाह है तो संभव है, मोदी है तो मुमकिन हैं | 0.038 |
| पुष्पेंद्र कुलश्रेष्ठ फैंस क्लब | 0.037 |
| Yogi Adityanath Fans Club | 0.036 |
| Sanatan Press | 0.035 |
| IndiaVirals | 0.034 |
| Tejasvi Surya Fans Club | 0.031 |

## Analysis of sub-graphs and communities

In order to better understand the interconnectivity between the actors and their coordinated behaviour, we next split the whole network into subgraphs and perform further analysis. The number of nodes and edges present in the top 10 biggest subgraphs (shown in Table 2) helps us to understand that most of the interactions within the network are circulated amongst the actors of the first sub-graph, which contains 412,914 nodes and 535,702 edges. There are a total of 54 subgraphs in our dataset, with the smallest subgraph containing 2 nodes and 1 edge. The pages with the highest degree centrality, closeness centrality and betweenness centrality corresponding to these individual sub-graphs are also shown in Table 2. In other words, the last column in Table 2 shows the most important or central actor in those sub-networks of information circulation.

**Table 2. Top 10 largest sub-graphs within the network and their characteristics**

| Sub-graph order | No. of nodes | No. of edges | Average distance between two nodes | Page with max centrality |
|---|---|---|---|---|
| 1 | 412,914 | 535,702 | 2.15 | *Pushpendra Kulshrestha Fans Club* |
| 2 | 633 | 632 | 2.00 | *Akhand_bharat* |
| 3 | 156 | 155 | 1.99 | *Hindu Heritage Endowment* |
| 4 | 155 | 154 | 1.99 | *হিন্দু রক্ষী দল, উধারবন্দ* |
| 5 | 118 | 117 | 1.98 | *APR News* |
| 6 | 35 | 34 | 1.94 | *United hindu front* |
| 7 | 34 | 33 | 1.94 | *Kattar Hindu group* |
| 8 | 33 | 32 | 1.94 | *लवजिहाद – प्रेम या जाल* |
| 9 | 30 | 29 | 1.93 | *Bajrang Dal Aryachauhan* |
| 10 | 25 | 24 | 1.92 | *Hindu Unity* |

In a connected graph, *closeness centrality* of a node is another measure of centrality in a network, calculated as the reciprocal of the sum of the length of the shortest paths between the node and all other nodes in the graph. Similarly, *betweenness centrality* is a measure that quantifies the number of shortest paths between pairs of nodes in the network that pass through a given node.

***Betweenness centrality*** *= No. of shortest paths through a node / No. of all possible shortest paths that exist between every pair of nodes in the graph*

In other words, a node with high betweenness centrality lies on many of the shortest paths connecting other pairs of nodes in the network. Mathematically speaking, betweenness centrality is defined as the fraction of the shortest paths in the network that pass through a given node. It is calculated by summing the number of shortest paths between all pairs of nodes in the network that pass through the node in question and then dividing by the total number of shortest paths between all pairs of nodes. Nodes with high betweenness centrality are often considered important 'connecting bridges' in the network, as they can control the flow of information between different parts of the network. Removing or disrupting nodes with high betweenness centrality can have a significant impact on the structure and function of the network. From our datasets, for all the top 10 sub-graphs, the page that is enlisted in Table 2 with the highest degree centrality, also has the maximum closeness centrality and betweenness centrality.

Subsequently, as a last step in our analysis, we have applied the Girvan-Newman algorithm to our dataset to detect communities within the network. The algorithm starts by calculating the betweenness centrality for all edges in the graph. Next, the edge with the highest betweenness centrality is removed from the graph. This process is repeated until the graph is split into multiple disconnected components. The resulting disconnected components that are thus found are the communities of the original graph. Out of 5, the top 3 communities that were identified by the algorithm are thereafter plotted by using Gephi visualisation software and shown in Figure 7-8.

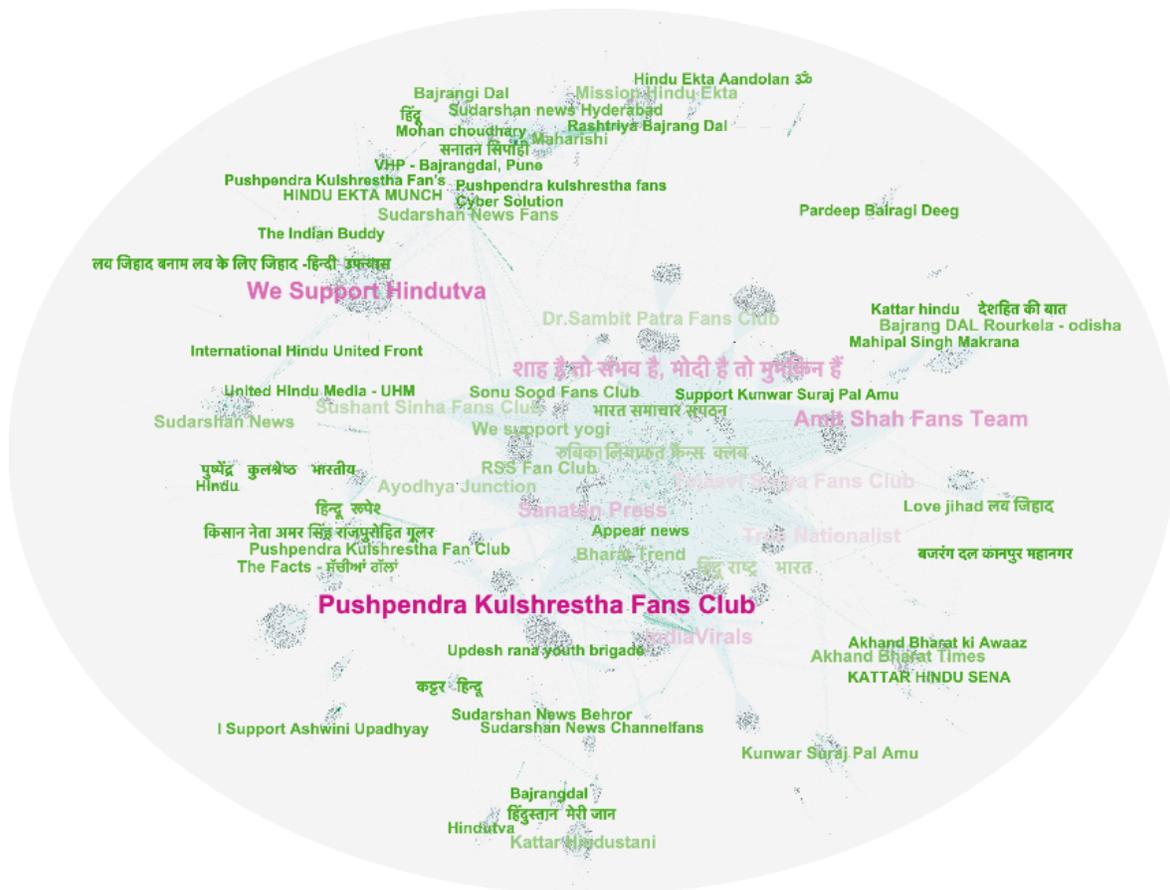

Figure 7. Gephi plot of the first community that was identified by the Girvan-Newman algorithm. The color map and the text size of page names correspond to the importance of the actors in terms of their activity.

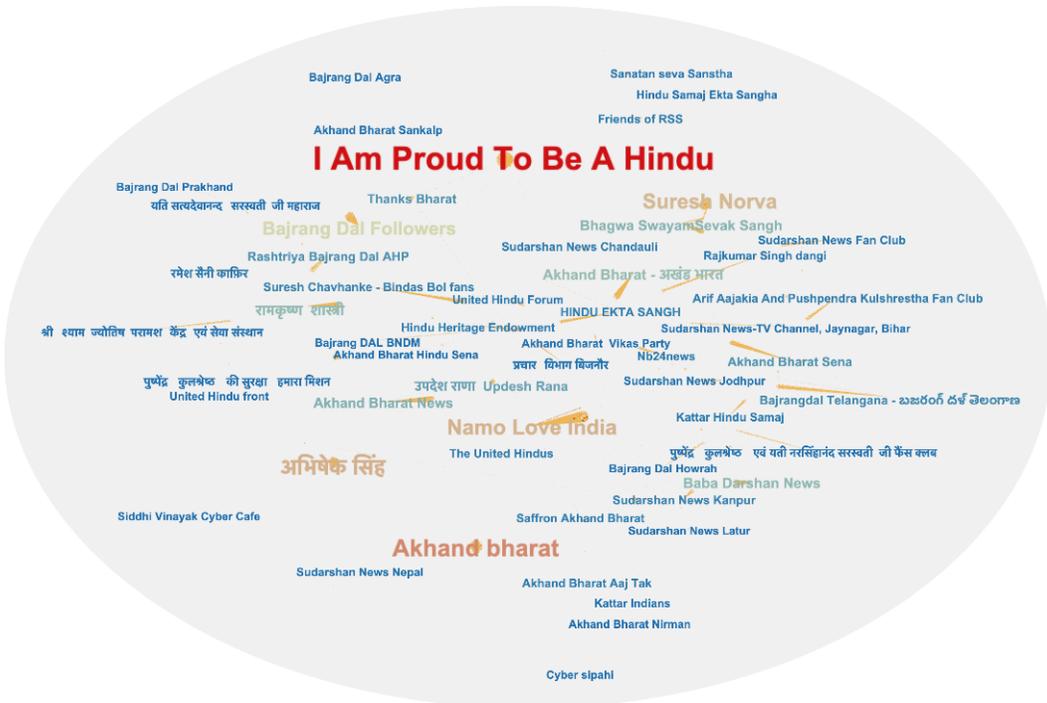

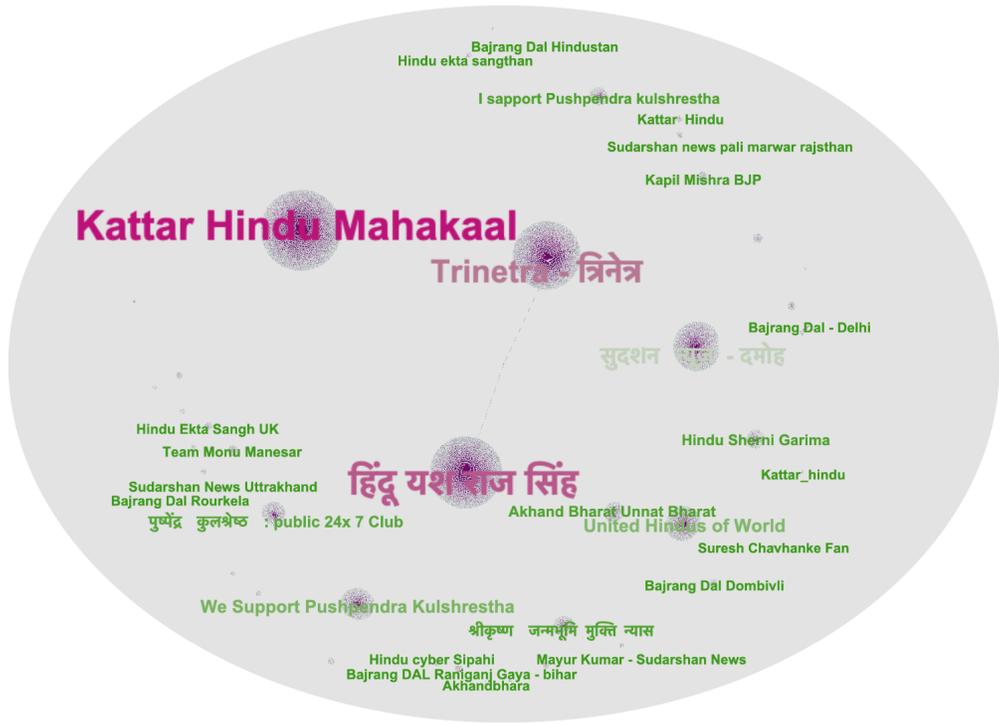

Figure 8. Gephi plots of the second (top) and third (bottom) communities that were identified by the Girvan-Newman algorithm. The color map and the text size of actor names correspond to the importance of the actors in terms of their activity.

# 5. Discussions and conclusions

The systematic data collection, reduction and network analysis method shown in this paper establishes a systematic scientific graph-theory based approach that can be adopted to discover communities and major actors in any type of social media data – be it Facebook, Twitter or Reddit. By analsying the primary actors within the first three communities (shown in Figure 7-8), we find that the top actors in these communities are *Pushpendra Kulshrestha Fans Club, I Am Proud To Be A Hindu* and *Kattar Hindu Mahakaal* respectively. The network of actors within these communities further expose their underlying causal coordinations in information dissemination process in Facebook. Although network analysis has been earlier applied to analyse the political narratives within the contexts of other countries, application of such novel community finding algorithm in the context of right-wing Hindutva discourse in India remains unaddressed in the literature. Such analysis thus pave the path forward for applying such novel network analysis algorithms to social media data, for automatically identifying thereatful coordinated communities, and to possibly model, predict and resist future threats emerging from coordinated information dissemination in the SMPs.

## Acknowledgements

The authors acknowledge the help received from the CrowdTangle – a Facebook-owned tool that tracks interactions on public content from Facebook pages and groups, verified profiles, Instagram accounts, and subreddits. The authors acknowledge the role of members of the *Foundation The London Story* in constructive discussions on this paper.

# Appendix 1. Description of CrowdTangle metadata entries and their types

| Metadata entry | Data type |
| --- | --- |
| account.name | String |
| account.handle | String |
| platformId | Integer |
| Page Category | String |
| Page Admin Top Country | String |
| Page Description | Text |
| Page Created | Date |
| subscriberCount | Integer |
| Followers at Posting | Integer |
| date | Date |
| Post Created Date | Date |
| Post Created Time | Time |
| type | String |
| totalInteraction | Integer |
| statistics.actual.likeCount | Integer |
| statistics.actual.commentCount | Integer |
| statistics.actual.shareCount | Integer |
| statistics.actual.loveCount | Integer |
| statistics.actual.wowCount | Integer |
| statistics.actual.hahaCount | Integer |
| statistics.actual.sadCount | Integer |
| statistics.actual.angryCount | Integer |
| statistics.actual.careCount | Integer |
| Video Share Status | String |
| Is Video Owner? | Binary |
| statistics.actual.videoPostViewCount | Integer |

| Metadata entry | Data type |
| --- | --- |
| statistics.actual.videoTotalViewCount | Integer |
| statistics.actual.videoAllCrosspostsViewCount | Integer |
| Video Length | String |
| postUrl | String |
| message | Text |
| expandedLinks.original | String |
| expandedLinks.expanded | String |
| imageText | Text |
| title | Text |
| description | Text |
| brandedContentSponsor.platformId | Integer |
| brandedContentSponsor.name | String |
| brandedContentSponsor.category | String |
| score | Float |

# Appendix 2: Top shared links, their counts and the corresponding page categories that have shared them

| Link | Count | Page Category | Count |
|---|---|---|---|
| https://sachkhabar.co.in/modi-governments-big-blow-to-zakir-naik/ | 49 | ACTIVITY_GENERAL | 9 |
| https://sachkhabar.co.in/modi-governments-big-blow-to-zakir-naik/ | 49 | NEWS_SITE | 8 |
| https://sachkhabar.co.in/modi-governments-big-blow-to-zakir-naik/ | 49 | MEDIA_NEWS_COMPANY | 6 |
| https://sachkhabar.co.in/modi-governments-big-blow-to-zakir-naik/ | 49 | PERSONAL_BLOG | 5 |
| https://sachkhabar.co.in/modi-governments-big-blow-to-zakir-naik/ | 49 | COMMUNITY | 3 |
| https://sachkhabar.co.in/modi-governments-big-blow-to-zakir-naik/ | 49 | POLITICIAN | 3 |
| https://sachkhabar.co.in/modi-governments-big-blow-to-zakir-naik/ | 49 | ART | 2 |
| https://sachkhabar.co.in/modi-governments-big-blow-to-zakir-naik/ | 49 | LOCAL | 2 |
| https://sachkhabar.co.in/modi-governments-big-blow-to-zakir-naik/ | 49 | NON_PROFIT | 2 |
| https://sachkhabar.co.in/modi-governments-big-blow-to-zakir-naik/ | 49 | POLITICAL_ORGANIZATION | 2 |
| https://sachkhabar.co.in/modi-governments-big-blow-to-zakir-naik/ | 49 | SOCIETY_SITE | 2 |
| https://sachkhabar.co.in/modi-governments-big-blow-to-zakir-naik/ | 49 | ACTOR | 1 |
| https://sachkhabar.co.in/modi-governments-big-blow-to-zakir-naik/ | 49 | FAN_PAGE | 1 |
| https://sachkhabar.co.in/modi-governments-big-blow-to-zakir-naik/ | 49 | NEWSAGENT_NEWSSTAND | 1 |
| https://sachkhabar.co.in/modi-governments-big-blow-to-zakir-naik/ | 49 | PERSON | 1 |
| https://sachkhabar.co.in/modi-governments-big-blow-to-zakir-naik/ | 49 | VIDEO_CREATOR | 1 |
| https://khabarbharattak.com/bobby-deols-son-gives-tough-competition-to-shahrukh-khans-son-aryan-khan-in-terms-of-smartness-see-photos/ | 47 | ACTIVITY_GENERAL | 8 |
| https://khabarbharattak.com/bobby-deols-son-gives-tough-competition-to-shahrukh-khans-son-aryan-khan-in-terms-of-smartness-see-photos/ | 47 | NEWS_SITE | 7 |
| https://khabarbharattak.com/bobby-deols-son-gives-tough-competition-to-shahrukh-khans-son-aryan-khan-in-terms-of-smartness-see-photos/ | 47 | MEDIA_NEWS_COMPANY | 6 |
| https://khabarbharattak.com/bobby-deols-son-gives-tough-competition-to-shahrukh-khans-son-aryan-khan-in-terms-of-smartness-see-photos/ | 47 | PERSONAL_BLOG | 6 |
| https://khabarbharattak.com/bobby-deols-son-gives-tough-competition-to-shahrukh-khans-son-aryan-khan-in-terms-of-smartness-see-photos/ | 47 | ART | 2 |
| https://khabarbharattak.com/bobby-deols-son-gives-tough-competition-to-shahrukh-khans-son-aryan-khan-in-terms-of-smartness-see-photos/ | 47 | COMMUNITY | 2 |
| https://khabarbharattak.com/bobby-deols-son-gives-tough-competition-to-shahrukh-khans-son-aryan-khan-in-terms-of-smartness-see-photos/ | 47 | FAN_PAGE | 2 |
| https://khabarbharattak.com/bobby-deols-son-gives-tough-competition-to-shahrukh-khans-son-aryan-khan-in-terms-of-smartness-see-photos/ | 47 | LOCAL | 2 |
| https://khabarbharattak.com/bobby-deols-son-gives-tough-competition-to-shahrukh-khans-son-aryan-khan-in-terms-of-smartness-see-photos/ | 47 | NON_PROFIT | 2 |
| https://khabarbharattak.com/bobby-deols-son-gives-tough-competition-to-shahrukh-khans-son-aryan-khan-in-terms-of-smartness-see-photos/ | 47 | POLITICAL_ORGANIZATION | 2 |
| https://khabarbharattak.com/bobby-deols-son-gives-tough-competition-to-shahrukh-khans-son-aryan-khan-in-terms-of-smartness-see-photos/ | 47 | POLITICIAN | 2 |

| Link | Count | Page Category | Count |
|---|---|---|---|
| https://khabarbharattak.com/bobby-deols-son-gives-tough-competition-to-shahrukh-khans-son-aryan-khan-in-terms-of-smartness-see-photos/ | 47 | SOCIETY_SITE | 2 |
| https://khabarbharattak.com/bobby-deols-son-gives-tough-competition-to-shahrukh-khans-son-aryan-khan-in-terms-of-smartness-see-photos/ | 47 | ACTOR | 1 |
| https://khabarbharattak.com/bobby-deols-son-gives-tough-competition-to-shahrukh-khans-son-aryan-khan-in-terms-of-smartness-see-photos/ | 47 | NEWSAGENT_NEWSSTAND | 1 |
| https://khabarbharattak.com/bobby-deols-son-gives-tough-competition-to-shahrukh-khans-son-aryan-khan-in-terms-of-smartness-see-photos/ | 47 | PERSON | 1 |
| https://khabarbharattak.com/bobby-deols-son-gives-tough-competition-to-shahrukh-khans-son-aryan-khan-in-terms-of-smartness-see-photos/ | 47 | VIDEO_CREATOR | 1 |
| https://sachkhabar.co.in/now-biden-wants-modis-help-immediately-only-india-can-save-the-world/ | 46 | ACTIVITY_GENERAL | 10 |
| https://sachkhabar.co.in/now-biden-wants-modis-help-immediately-only-india-can-save-the-world/ | 46 | NEWS_SITE | 6 |
| https://sachkhabar.co.in/now-biden-wants-modis-help-immediately-only-india-can-save-the-world/ | 46 | POLITICIAN | 5 |
| https://sachkhabar.co.in/now-biden-wants-modis-help-immediately-only-india-can-save-the-world/ | 46 | PERSONAL_BLOG | 3 |
| https://sachkhabar.co.in/now-biden-wants-modis-help-immediately-only-india-can-save-the-world/ | 46 | ACTOR | 2 |
| https://sachkhabar.co.in/now-biden-wants-modis-help-immediately-only-india-can-save-the-world/ | 46 | ART | 2 |
| https://sachkhabar.co.in/now-biden-wants-modis-help-immediately-only-india-can-save-the-world/ | 46 | COMMUNITY | 2 |
| https://sachkhabar.co.in/now-biden-wants-modis-help-immediately-only-india-can-save-the-world/ | 46 | FAN_PAGE | 2 |
| https://sachkhabar.co.in/now-biden-wants-modis-help-immediately-only-india-can-save-the-world/ | 46 | LOCAL | 2 |
| https://sachkhabar.co.in/now-biden-wants-modis-help-immediately-only-india-can-save-the-world/ | 46 | MEDIA_NEWS_COMPANY | 2 |
| https://sachkhabar.co.in/now-biden-wants-modis-help-immediately-only-india-can-save-the-world/ | 46 | NEWSAGENT_NEWSSTAND | 2 |
| https://sachkhabar.co.in/now-biden-wants-modis-help-immediately-only-india-can-save-the-world/ | 46 | NON_PROFIT | 2 |
| https://sachkhabar.co.in/now-biden-wants-modis-help-immediately-only-india-can-save-the-world/ | 46 | POLITICAL_ORGANIZATION | 2 |
| https://sachkhabar.co.in/now-biden-wants-modis-help-immediately-only-india-can-save-the-world/ | 46 | SOCIETY_SITE | 2 |
| https://sachkhabar.co.in/now-biden-wants-modis-help-immediately-only-india-can-save-the-world/ | 46 | VIDEO_CREATOR | 2 |
| https://khabarbharattak.com/sapna-choudharys-sons-name-is-porus/ | 42 | ACTIVITY_GENERAL | 9 |
| https://khabarbharattak.com/sapna-choudharys-sons-name-is-porus/ | 42 | MEDIA_NEWS_COMPANY | 5 |
| https://khabarbharattak.com/sapna-choudharys-sons-name-is-porus/ | 42 | NEWS_SITE | 5 |
| https://khabarbharattak.com/sapna-choudharys-sons-name-is-porus/ | 42 | PERSONAL_BLOG | 4 |
| https://khabarbharattak.com/sapna-choudharys-sons-name-is-porus/ | 42 | POLITICIAN | 3 |

| Link | Count | Page Category | Count |
|---|---|---|---|
| **https://khabarbharattak.com/sapna-choudharys-sons-name-is-porus/** | 42 | ART | 2 |
| **https://khabarbharattak.com/sapna-choudharys-sons-name-is-porus/** | 42 | FAN_PAGE | 2 |
| **https://khabarbharattak.com/sapna-choudharys-sons-name-is-porus/** | 42 | LOCAL | 2 |
| **https://khabarbharattak.com/sapna-choudharys-sons-name-is-porus/** | 42 | NON_PROFIT | 2 |
| **https://khabarbharattak.com/sapna-choudharys-sons-name-is-porus/** | 42 | POLITICAL_ORGANIZATION | 2 |
| **https://khabarbharattak.com/sapna-choudharys-sons-name-is-porus/** | 42 | SOCIETY_SITE | 2 |
| **https://khabarbharattak.com/sapna-choudharys-sons-name-is-porus/** | 42 | ACTOR | 1 |
| **https://khabarbharattak.com/sapna-choudharys-sons-name-is-porus/** | 42 | COMMUNITY | 1 |
| **https://khabarbharattak.com/sapna-choudharys-sons-name-is-porus/** | 42 | NEWSAGENT_NEWSSTAND | 1 |
| **https://khabarbharattak.com/sapna-choudharys-sons-name-is-porus/** | 42 | VIDEO_CREATOR | 1 |
| **https://appearnews.com/social-media-bride-video/** | 41 | ACTIVITY_GENERAL | 8 |
| **https://appearnews.com/social-media-bride-video/** | 41 | POLITICIAN | 6 |
| **https://appearnews.com/social-media-bride-video/** | 41 | ART | 4 |
| **https://appearnews.com/social-media-bride-video/** | 41 | NEWS_SITE | 4 |
| **https://appearnews.com/social-media-bride-video/** | 41 | MEDIA_NEWS_COMPANY | 3 |
| **https://appearnews.com/social-media-bride-video/** | 41 | ACTOR | 2 |
| **https://appearnews.com/social-media-bride-video/** | 41 | COMMUNITY | 2 |
| **https://appearnews.com/social-media-bride-video/** | 41 | FAN_PAGE | 2 |
| **https://appearnews.com/social-media-bride-video/** | 41 | LOCAL | 2 |
| **https://appearnews.com/social-media-bride-video/** | 41 | NEWSAGENT_NEWSSTAND | 2 |
| **https://appearnews.com/social-media-bride-video/** | 41 | PERSONAL_BLOG | 2 |
| **https://appearnews.com/social-media-bride-video/** | 41 | POLITICAL_ORGANIZATION | 2 |
| **https://appearnews.com/social-media-bride-video/** | 41 | VIDEO_CREATOR | 2 |
| **https://khabarbharattak.com/know-how-govinda-became-superstar-sadhvi-mother-became-muslim-after-birth-father-had-refused-to-adopt/** | 41 | ACTIVITY_GENERAL | 7 |
| **https://khabarbharattak.com/know-how-govinda-became-superstar-sadhvi-mother-became-muslim-after-birth-father-had-refused-to-adopt/** | 41 | PERSONAL_BLOG | 6 |
| **https://khabarbharattak.com/know-how-govinda-became-superstar-sadhvi-mother-became-muslim-after-birth-father-had-refused-to-adopt/** | 41 | MEDIA_NEWS_COMPANY | 5 |
| **https://khabarbharattak.com/know-how-govinda-became-superstar-sadhvi-mother-became-muslim-after-birth-father-had-refused-to-adopt/** | 41 | NEWS_SITE | 4 |
| **https://khabarbharattak.com/know-how-govinda-became-superstar-sadhvi-mother-became-muslim-after-birth-father-had-refused-to-adopt/** | 41 | ART | 3 |
| **https://khabarbharattak.com/know-how-govinda-became-superstar-sadhvi-mother-became-muslim-after-birth-father-had-refused-to-adopt/** | 41 | POLITICIAN | 3 |
| **https://khabarbharattak.com/know-how-govinda-became-superstar-sadhvi-mother-became-muslim-after-birth-father-had-refused-to-adopt/** | 41 | COMMUNITY | 2 |

| Link | Count | Page Category | Count |
|---|---|---|---|
| https://khabarbharattak.com/know-how-govinda-became-superstar-sadhvi-mother-became-muslim-after-birth-father-had-refused-to-adopt/ | 41 | FAN_PAGE | 2 |
| https://khabarbharattak.com/know-how-govinda-became-superstar-sadhvi-mother-became-muslim-after-birth-father-had-refused-to-adopt/ | 41 | LOCAL | 2 |
| https://khabarbharattak.com/know-how-govinda-became-superstar-sadhvi-mother-became-muslim-after-birth-father-had-refused-to-adopt/ | 41 | ACTOR | 1 |
| https://khabarbharattak.com/know-how-govinda-became-superstar-sadhvi-mother-became-muslim-after-birth-father-had-refused-to-adopt/ | 41 | NEWSAGENT_NEWSSTAND | 1 |
| https://khabarbharattak.com/know-how-govinda-became-superstar-sadhvi-mother-became-muslim-after-birth-father-had-refused-to-adopt/ | 41 | NON_PROFIT | 1 |
| https://khabarbharattak.com/know-how-govinda-became-superstar-sadhvi-mother-became-muslim-after-birth-father-had-refused-to-adopt/ | 41 | PERSON | 1 |
| https://khabarbharattak.com/know-how-govinda-became-superstar-sadhvi-mother-became-muslim-after-birth-father-had-refused-to-adopt/ | 41 | POLITICAL_ORGANIZATION | 1 |
| https://khabarbharattak.com/know-how-govinda-became-superstar-sadhvi-mother-became-muslim-after-birth-father-had-refused-to-adopt/ | 41 | SOCIETY_SITE | 1 |
| https://khabarbharattak.com/know-how-govinda-became-superstar-sadhvi-mother-became-muslim-after-birth-father-had-refused-to-adopt/ | 41 | VIDEO_CREATOR | 1 |
| https://appearnews.com/7-jokes/ | 40 | ACTIVITY_GENERAL | 8 |
| https://appearnews.com/7-jokes/ | 40 | POLITICIAN | 6 |
| https://appearnews.com/7-jokes/ | 40 | NEWS_SITE | 4 |
| https://appearnews.com/7-jokes/ | 40 | ACTOR | 3 |
| https://appearnews.com/7-jokes/ | 40 | ART | 3 |
| https://appearnews.com/7-jokes/ | 40 | COMMUNITY | 3 |
| https://appearnews.com/7-jokes/ | 40 | FAN_PAGE | 2 |
| https://appearnews.com/7-jokes/ | 40 | LOCAL | 2 |
| https://appearnews.com/7-jokes/ | 40 | MEDIA_NEWS_COMPANY | 2 |
| https://appearnews.com/7-jokes/ | 40 | NEWSAGENT_NEWSSTAND | 2 |
| https://appearnews.com/7-jokes/ | 40 | POLITICAL_ORGANIZATION | 2 |
| https://appearnews.com/7-jokes/ | 40 | VIDEO_CREATOR | 2 |
| https://appearnews.com/7-jokes/ | 40 | PERSONAL_BLOG | 1 |
| https://khabarbharattak.com/will-rohit-sardana-be-honored-with-the-padma-shri-award/ | 39 | ACTIVITY_GENERAL | 8 |
| https://khabarbharattak.com/will-rohit-sardana-be-honored-with-the-padma-shri-award/ | 39 | NEWS_SITE | 5 |
| https://khabarbharattak.com/will-rohit-sardana-be-honored-with-the-padma-shri-award/ | 39 | PERSONAL_BLOG | 4 |
| https://khabarbharattak.com/will-rohit-sardana-be-honored-with-the-padma-shri-award/ | 39 | MEDIA_NEWS_COMPANY | 3 |
| https://khabarbharattak.com/will-rohit-sardana-be-honored-with-the-padma-shri-award/ | 39 | NEWSAGENT_NEWSSTAND | 3 |

| Link | Count | Page Category | Count |
|---|---|---|---|
| https://khabarbharattak.com/will-rohit-sardana-be-honored-with-the-padma-shri-award/ | 39 | ACTOR | 2 |
| https://khabarbharattak.com/will-rohit-sardana-be-honored-with-the-padma-shri-award/ | 39 | COMMUNITY | 2 |
| https://khabarbharattak.com/will-rohit-sardana-be-honored-with-the-padma-shri-award/ | 39 | FAN_PAGE | 2 |
| https://khabarbharattak.com/will-rohit-sardana-be-honored-with-the-padma-shri-award/ | 39 | POLITICAL_ORGANIZATION | 2 |
| https://khabarbharattak.com/will-rohit-sardana-be-honored-with-the-padma-shri-award/ | 39 | POLITICIAN | 2 |
| https://khabarbharattak.com/will-rohit-sardana-be-honored-with-the-padma-shri-award/ | 39 | ART | 1 |
| https://khabarbharattak.com/will-rohit-sardana-be-honored-with-the-padma-shri-award/ | 39 | LOCAL | 1 |
| https://khabarbharattak.com/will-rohit-sardana-be-honored-with-the-padma-shri-award/ | 39 | NON_PROFIT | 1 |
| https://khabarbharattak.com/will-rohit-sardana-be-honored-with-the-padma-shri-award/ | 39 | PERSON | 1 |
| https://khabarbharattak.com/will-rohit-sardana-be-honored-with-the-padma-shri-award/ | 39 | SOCIETY_SITE | 1 |
| https://khabarbharattak.com/will-rohit-sardana-be-honored-with-the-padma-shri-award/ | 39 | VIDEO_CREATOR | 1 |
| https://khabarbharattak.com/149-year-olds-darbar-move-tradition-ended-in-jammu-and-kashmir-due-to-pm-modi-and-amit-shah/ | 38 | POLITICIAN | 7 |
| https://khabarbharattak.com/149-year-olds-darbar-move-tradition-ended-in-jammu-and-kashmir-due-to-pm-modi-and-amit-shah/ | 38 | ACTIVITY_GENERAL | 6 |
| https://khabarbharattak.com/149-year-olds-darbar-move-tradition-ended-in-jammu-and-kashmir-due-to-pm-modi-and-amit-shah/ | 38 | NEWS_SITE | 4 |
| https://khabarbharattak.com/149-year-olds-darbar-move-tradition-ended-in-jammu-and-kashmir-due-to-pm-modi-and-amit-shah/ | 38 | MEDIA_NEWS_COMPANY | 3 |
| https://khabarbharattak.com/149-year-olds-darbar-move-tradition-ended-in-jammu-and-kashmir-due-to-pm-modi-and-amit-shah/ | 38 | ACTOR | 2 |
| https://khabarbharattak.com/149-year-olds-darbar-move-tradition-ended-in-jammu-and-kashmir-due-to-pm-modi-and-amit-shah/ | 38 | ART | 2 |
| https://khabarbharattak.com/149-year-olds-darbar-move-tradition-ended-in-jammu-and-kashmir-due-to-pm-modi-and-amit-shah/ | 38 | COMMUNITY | 2 |
| https://khabarbharattak.com/149-year-olds-darbar-move-tradition-ended-in-jammu-and-kashmir-due-to-pm-modi-and-amit-shah/ | 38 | FAN_PAGE | 2 |
| https://khabarbharattak.com/149-year-olds-darbar-move-tradition-ended-in-jammu-and-kashmir-due-to-pm-modi-and-amit-shah/ | 38 | LOCAL | 2 |
| https://khabarbharattak.com/149-year-olds-darbar-move-tradition-ended-in-jammu-and-kashmir-due-to-pm-modi-and-amit-shah/ | 38 | NEWSAGENT_NEWSSTAND | 2 |
| https://khabarbharattak.com/149-year-olds-darbar-move-tradition-ended-in-jammu-and-kashmir-due-to-pm-modi-and-amit-shah/ | 38 | PERSONAL_BLOG | 2 |
| https://khabarbharattak.com/149-year-olds-darbar-move-tradition-ended-in-jammu-and-kashmir-due-to-pm-modi-and-amit-shah/ | 38 | POLITICAL_ORGANIZATION | 2 |

| Link | Count | Page Category | Count |
|---|---|---|---|
| https://khabarbharattak.com/149-year-olds-darbar-move-tradition-ended-in-jammu-and-kashmir-due-to-pm-modi-and-amit-shah/ | 38 | VIDEO_CREATOR | 2 |
| https://khabarbharattak.com/these-bollywood-stars-whose-marriage-was-broken-in-a-few-years-some-for-9-months-and-some-for-2-years-see-the-full-list/ | 38 | ACTIVITY_GENERAL | 6 |
| https://khabarbharattak.com/these-bollywood-stars-whose-marriage-was-broken-in-a-few-years-some-for-9-months-and-some-for-2-years-see-the-full-list/ | 38 | NEWS_SITE | 6 |
| https://khabarbharattak.com/these-bollywood-stars-whose-marriage-was-broken-in-a-few-years-some-for-9-months-and-some-for-2-years-see-the-full-list/ | 38 | MEDIA_NEWS_COMPANY | 4 |
| https://khabarbharattak.com/these-bollywood-stars-whose-marriage-was-broken-in-a-few-years-some-for-9-months-and-some-for-2-years-see-the-full-list/ | 38 | PERSONAL_BLOG | 4 |
| https://khabarbharattak.com/these-bollywood-stars-whose-marriage-was-broken-in-a-few-years-some-for-9-months-and-some-for-2-years-see-the-full-list/ | 38 | POLITICIAN | 3 |
| https://khabarbharattak.com/these-bollywood-stars-whose-marriage-was-broken-in-a-few-years-some-for-9-months-and-some-for-2-years-see-the-full-list/ | 38 | COMMUNITY | 2 |
| https://khabarbharattak.com/these-bollywood-stars-whose-marriage-was-broken-in-a-few-years-some-for-9-months-and-some-for-2-years-see-the-full-list/ | 38 | NEWSAGENT_NEWSSTAND | 2 |
| https://khabarbharattak.com/these-bollywood-stars-whose-marriage-was-broken-in-a-few-years-some-for-9-months-and-some-for-2-years-see-the-full-list/ | 38 | NON_PROFIT | 2 |
| https://khabarbharattak.com/these-bollywood-stars-whose-marriage-was-broken-in-a-few-years-some-for-9-months-and-some-for-2-years-see-the-full-list/ | 38 | SOCIETY_SITE | 2 |
| https://khabarbharattak.com/these-bollywood-stars-whose-marriage-was-broken-in-a-few-years-some-for-9-months-and-some-for-2-years-see-the-full-list/ | 38 | ACTOR | 1 |
| https://khabarbharattak.com/these-bollywood-stars-whose-marriage-was-broken-in-a-few-years-some-for-9-months-and-some-for-2-years-see-the-full-list/ | 38 | ART | 1 |
| https://khabarbharattak.com/these-bollywood-stars-whose-marriage-was-broken-in-a-few-years-some-for-9-months-and-some-for-2-years-see-the-full-list/ | 38 | FAN_PAGE | 1 |
| https://khabarbharattak.com/these-bollywood-stars-whose-marriage-was-broken-in-a-few-years-some-for-9-months-and-some-for-2-years-see-the-full-list/ | 38 | LOCAL | 1 |
| https://khabarbharattak.com/these-bollywood-stars-whose-marriage-was-broken-in-a-few-years-some-for-9-months-and-some-for-2-years-see-the-full-list/ | 38 | PERSON | 1 |
| https://khabarbharattak.com/these-bollywood-stars-whose-marriage-was-broken-in-a-few-years-some-for-9-months-and-some-for-2-years-see-the-full-list/ | 38 | POLITICAL_ORGANIZATION | 1 |
| https://khabarbharattak.com/these-bollywood-stars-whose-marriage-was-broken-in-a-few-years-some-for-9-months-and-some-for-2-years-see-the-full-list/ | 38 | VIDEO_CREATOR | 1 |